\documentclass[twocolumn,amssymb,superscriptaddress,longbibliography]{revtex4-1}
\usepackage{graphicx}
\usepackage{amsfonts}
\usepackage{amsmath}
\usepackage{bm}
\usepackage{amssymb}
\usepackage{braket}

\usepackage{multirow}
\usepackage{epstopdf}
\usepackage[normalem]{ulem}

\usepackage[monochrome]{color}

\usepackage{hyperref}
\hypersetup{
colorlinks=true,
linkcolor=blue,
citecolor=blue,
filecolor=green,
urlcolor=blue,
}

\newcommand{\threejm}[6]{ \left(\begin{array}{ccc} #1 & #3 & #5\\
                                              #2 & #4 & #6
                                \end{array}
                          \right)}
\newcommand{\sixj}[6]{ \left\{\begin{array}{ccc} #1 & #3 & #5\\
                                              #2 & #4 & #6
                                \end{array}
                          \right\}}

\makeatletter
\def\@bibdataout@aps{%
\immediate\write\@bibdataout{%
@CONTROL{%
apsrev41Control%
\longbibliography@sw{%
    ,author="08",editor="1",pages="1",title="0",year="1"%
    }{%
    ,author="08",editor="1",pages="1",title="",year="1"%
    }%
  }%
}%
\if@filesw \immediate \write \@auxout {\string \citation {apsrev41Control}}\fi 
}
\makeatother

\begin{document}
\title{Ultracold molecular collisions in magnetic fields: Efficient incorporation of hyperfine structure in the total rotational angular momentum representation}

\author{Timur V. Tscherbul}
\affiliation{Department of Physics, University of Nevada, Reno, Nevada, 89557, USA}
\author{Jose P. D'Incao}
\affiliation{JILA, National Institute of Standards and Technology, and Department of Physics, University of Colorado,  Boulder, Colorado, 80309, USA}

\date{\today}
\begin{abstract}
The effects of hyperfine structure  on ultracold molecular collisions in external fields are largely unexplored due to major computational challenges associated with  rapidly proliferating hyperfine and rotational channels coupled by highly anisotropic intermolecular interactions.
We explore a new basis set for incorporating the  effects of hyperfine structure and external magnetic fields in quantum scattering calculations on ultracold molecular collisions.  The basis is composed of direct products of  the eigenfunctions of the total {\it rotational} angular momentum (TRAM) of the collision complex $J_r$ 
 and the electron/nuclear spin basis functions of the collision partners.
  The separation of the rotational and spin degrees of freedom 
   ensures rigorous conservation of $J_r$ even in the presence of external magnetic fields and isotropic hyperfine  interactions. 
   The resulting block-diagonal structure of   the scattering Hamiltonian  enables coupled-channel calculations on highly anisotropic  atom-molecule and molecule-molecule collisions to be performed independently for each value of $J_r$.
    We illustrate the efficiency of the TRAM basis by calculating state-to-state cross sections for ultracold He~+~YbF collisions in a magnetic field. The size of the TRAM basis required to reach numerical convergence is 8 times smaller than that of the uncoupled basis used previously, providing a computational gain of three orders of magnitude.
    The TRAM basis  is therefore well suited for rigorous quantum scattering calculations on ultracold molecular collisions in the presence of hyperfine interactions and external magnetic fields.
\end{abstract}

\maketitle

\newpage

\section{Introduction}

Ultracold molecular gases offer novel opportunities for searches of new physics beyond the Standard Model \cite{Carr:09,Bohn:17}, quantum information science \cite{Yelin:06,Gorshkov:11, Albert:20}, and quantum control of chemical reaction dynamics \cite{Krems:08,Balakrishnan:16,Krems:19,Liu:22}.
Understanding the quantum dynamics of ultracold molecular collisions is essential to realizing these opportunities by enabling the production of denser and cold molecular ensembles,  and to controlling intermolecular interactions within the ensembles. This is because collisional properties determine the properties of ultracold molecular gases, such as their stability. In particular, inelastic collisions and long-lived complex formation lead to trap loss \cite{Mayle:13,Liu:20}, which limits the lifetime of trapped molecules, whereas elastic collisions result in thermalization, which is beneficial for evaporative and sympathetic cooling \cite{Carr:09,Bohn:17,Tscherbul:11,Morita:18,Morita:17}. 

The most detailed theoretical understanding of complex molecular collisions is gained from rigorous quantum coupled-channel (CC) calculations, which solve the Schr\"odinger equation exactly for a given form of the interaction potential between the molecules \cite{Child:74,Althorpe:03}. Because they involve no approximations, these calculations can be used to interpret experimental observations and to relate them to the underlying microscopic interactions between the molecules \cite{Krems:03,Balakrishnan:03,Campbell:09,Hummon:11,Chefdeville:13,Vogels:15,Bergeat:15,Klein:17,Vogels:18,Jongh:20}. These calculations also serve as a benchmark for approximate methods and can be used to estimate inelastic collision rates \cite{Morita:19b} and  the density of scattering resonances in ultracold  atom-molecule  \cite{Tscherbul:09c,Tscherbul:11,Morita:18,Tscherbul:15,Morita:20} and molecule-molecule  \cite{Tscherbul:09c}  collisions.

However, acquiring such a detailed understanding of ultracold molecular collisions has been a major challenge due to the need to account for the numerous molecular degrees of freedom, which include  rotational, vibrational, fine, and hyperfine structure, in addition to the interaction with external electromagnetic fields. In order to obtain numerically converged solutions of CC equations, one needs to use a very large number of molecular basis functions, which grows rapidly with the size and mass  of the colliding molecules \cite{Truhlar:94}. 
 While efficient techniques have been developed  for reducing the size of rotational basis sets in the absence  \cite{Arthurs:60,Truhlar:94} and in the presence  \cite{Tscherbul:10,Tscherbul:12,Suleimanov:12,Koyu:22} of  external electromagnetic fields, there has been very little work on hyperfine basis sets.

Molecular hyperfine structure arises due to nonzero nuclear spins in one (or both) of the collision partners interacting with unpaired electrons and/or with molecular rotation \cite{Brown:03}. These interactions result in energy level splittings on the order of a few tens of kHz \cite{Aldegunde:08} to GHz \cite{Collopy:15,Burau:23}, which can easily exceed the energy scale of ultracold molecular collisions ($E\le100$~kHz). Thus, hyperfine interactions are expected to profoundly affect  ultracold collision dynamics by shifting and splitting  collision thresholds and modifying zero-energy crossings of molecular bound states, which give rise to magnetic Feshbach resonances \cite{Chin:10}.  Indeed, with a few exceptions (such as collisions of spin-polarized species in strong magnetic fields \cite{Tscherbul:11,Morita:17,Morita:18}), it is impossible to accurately describe the magnetic field dependence of scattering cross sections at ultralow temperatures without taking into account the hyperfine structure. The lifetimes of collision complexes  may be affected by the hyperfine structure \cite{Mayle:12,Mayle:13} and strong effects of hyperfine interactions have been observed in product state distributions of the ultracold chemical reaction KRb~+~KRb $\to$ K$_2$~+~Rb$_2$ \cite{Hu:21,Quemener:21}.
 Many recent experimental studies of ultracold atom-molecule \cite{Yang:19,Zhen:22,Nichols:22,Park:23b,Karman:23}  and molecule-molecule \cite{Hu:19,Liu:22,Park:23} collisions involve  alkali-metal dimer molecules with a pronounced hyperfine structure.  

These considerations strongly motivate including hyperfine interactions in rigorous CC calculations  on molecular collisions in external fields. However, most of the previous calculations have neglected these interactions, as their inclusion leads to formidable computational difficulties associated with the rapid expansion of the Hilbert space. A single magnetic nucleus with spin $I$ gives rise to a $(2I+1)$-fold increase in the number of molecular basis states, leading to a several-fold increase in the number of scattering channels.
For example, in the case of cold He~+~YbF collisions considered in this work, nearly converged results can be obtained with 1131   channels in the fully uncoupled basis \cite{Volpi:02,Krems:04,Tscherbul:07} in the absence of hyperfine structure.
Incorporating the hyperfine structure of YbF rises this number by a factor of two, and the computational cost by a factor of 8. 
It is therefore not surprising why so few CC calculations on ultracold atom-molecule collisions have included hyperfine interactions.  These calculations are reviewed below.


Lara {\it et al.} included hyperfine structure in their scattering calculations of ultracold Rb~+~OH collisions in the absence of external fields \cite{Lara:06,Lara:07}. \textcolor{red}{Tscherbul, Krems, and co-workers} performed quantum scattering calculations on  He~+~YbF  collisions in a magnetic field \cite{Tscherbul:07}, which highlighted the importance of including hyperfine structure in CC calculations of cold and ultracold atom-molecule scattering. 
This work has been extended to collisions of $^3\Sigma$ molecules with atoms  by Gonz\'alez-Mart\'inez and Hutson \cite{GonzalezMartinez:11}. Nuclear spin relaxation in weakly anisotropic He~+~$^{13}$CO collisions has recently been explored by Hermsmeier {\it et al.} in converged CC calculations in the temperature range 0.1-10 K \cite{Hermsmeier:23}.
Hyperfine structure was included in model CC  calculations on ultracold collisions of RbCs molecules \cite{Wallis:14} and on the chemical reactions Li~+~CaH $\to$ LiH + Ca \cite{Tscherbul:20} and Na~+~NaLi $\to$ Na$_2$~+~Li  \cite{Hermsmeier:21,Karman:23} in the presence of an external magnetic field. These calculations, however, used restricted CC basis sets containing only the lowest rotational states  \cite{Hermsmeier:21,Karman:23},
and did not produce converged results when hyperfine interactions were included.

The vast majority of the previous  CC calculations that included molecular hyperfine structure used ``uncoupled'' channel basis sets of the form $\ket{f_\text{mol}}\ket{lm_l}$, where $\ket{f_\text{mol}}$ are the basis functions that depend on the molecular (internal) degrees of freedom and $\ket{lm_l}$ are the partial wave basis states, which are the eigenfunctions of the orbital angular momentum squared of the collision complex $\hat{l}^2$ and of its space-fixed projection $\hat{l}_Z$. Because these basis states are not the eigenfunctions of the total angular momentum of the collision complex, they could provide converged results only for moderately anisotropic systems, such as He~+~YbF and Mg~+~NH, where only a few rotational states and partial waves are necessary to properly describe  the anisotropy of the atom-molecule interaction potential. However, with the exception of collisions involving light atoms (such as He, He$^*$, Li) and molecules (such as H$_2$, CaH, NH, and  O$_2$) \cite{Krems:03,Balakrishnan:03,Campbell:09,Hummon:11,Vogels:15,Bergeat:15,Klein:17,Akerman:17},  ultracold atom-molecule collisions studied experimentally thus far (such as Rb~+~CaF \cite{Jurgilas:21},  Na~+~NaLi \cite{Son:20,Son:22}, K~+~NaK \cite{Yang:19,Zhen:22}, and Rb~+~KRb \cite{Nichols:22})  are characterized by deep and strongly anisotropic interactions, which  couple hundreds of rotational states at short range \cite{Morita:17,Morita:18}.

Obtaining converged results for such collisions requires the use of the total angular momentum  (TAM)  representation \cite{Arthurs:60}, which leverages the rotational invariance of intermolecular interactions to block-diagonalize the scattering Hamiltonian in the absence of external fields. The TAM representation has been widely used to study
ultracold atom-molecule and molecule-molecule collisions and chemical reactions under field-free conditions \cite{Balakrishnan:16,Kendrick:21,Morita:23} and in the presence of external electromagnetic fields  \cite{Tscherbul:10,Tscherbul:12,Suleimanov:12,Tscherbul:11,Tscherbul:15,Morita:18,Koyu:22}. However,  the hyperfine effects are yet to be incorporated in TAM calculations in the presence of external fields.

Here, we explore an alternative basis set that  combines the computational efficiency of the TAM basis with the ease of evaluation of matrix elements  pertinent to the fully uncoupled basis. The basis is obtained by coupling all {\it rotational} angular momenta in the Hamiltonian to form  the total rotational angular momentum (TRAM) of the collision complex.  
To our knowledge, the  TRAM basis was first used by Simoni and Launay \cite{Simoni:06} in their model CC calculations of ultracold Na~+~Na$_2$ collisions. More recently, similar  treatments have been used for ultracold three-atom recombination reactions \cite{Chapurin:19,Xie:20,DIncao:23}.
 However, the calculations of Ref.~\cite{Simoni:06} were performed in the absence of external fields and did not include the intramolecular spin-rotation and anisotropic hyperfine interactions, which are generally non-negligible in ultracold atom-molecule collisions \cite{Tscherbul:07,GonzalezMartinez:11}. In addition,  the CC basis sets employed in  Ref.~\cite{Simoni:06} were too small to produce converged results, leaving the question open of whether the TRAM basis could be used for efficient CC computations on ultracold molecular collisions in magnetic  fields.
Here, we  address this question by systematically considering all intramolecular interactions in $^2\Sigma$ molecules and  performing converged CC calculations on ultracold He~+~YbF collisions in an external magnetic field. Our results show that the TRAM basis offers a computationally efficient way of handling hyperfine interactions in ultracold  atom-molecule  collisions mediated by strongly anisotropic interactions. Additional advantages of the TRAM basis include \textcolor{red}{(i) the absence of unphysical states 
in the ground rotational manifold of the diatomic molecule and their low density in rotationally excited manifolds}, and (ii)
 its superior computational efficiency over the TAM basis in situations, where ultracold scattering is dominated by isotropic hyperfine and Zeeman interactions.
These results open up the possibility of rigorous quantum scattering calculations on ultracold atom-molecule collisions of current experimental interest \cite{Son:22,Yang:19,Nichols:22}.


The structure of this paper is  as follows.  In Sec. II we define  the TRAM basis set and highlight its computational advantages.
  In Sec. III we apply the theory to calculate the cross sections for  cold He~+~YbF collisions, a system with a pronounced hyperfine structure, which presented significant computational challenges in a previous theoretical study using an uncoupled space-fixed basis set \cite{Tscherbul:07}. We show how these challenges can be efficiently overcome using the TRAM basis set, leading to an order of magnitude reduction in the number of coupled channels (from 1920 to 240), which translates to a nearly three orders of magnitude reduction in computational effort.  Section IV concludes and outlines several possible directions for future work.

\section{Theory}

In this section, we  define the TRAM basis set and present expressions for the matrix elements of the scattering Hamiltonian in the TRAM representation. We then discuss  several computational advantages of the TRAM basis, which make it an attractive choice for quantum scattering calculations on strongly anisotropic molecular collisions in an external magnetic field in the presence of hyperfine structure.  We will consider two cases of interest:  collisions of $^2\Sigma$ molecules with structureless atoms, and collisions of $^2\Sigma$  molecules with atoms in electronic states of $^2$S symmetry. Open-shell $^2\Sigma$ molecules such as SrF,  CaF, or YbF have recently been laser cooled and trapped by a number of research groups \cite{Barry:14,McCarron:18,Anderegg:18,Lim:18,Langin:21,Langin:23}. 
Ultracold atom-molecule collisions are relevant for sympathetic cooling, in which molecules  thermalize with an ultracold gas of atoms \cite{Hutzler:12,Tscherbul:11,Morita:17,Morita:18}. Recently, several experimental groups  have measured  the  cross sections for ultracold Rb~+~CaF  \cite{Jurgilas:21} and Na~+~NaLi collisions    \cite{Son:20,Son:22,Park:23b} and  observed magnetic Feshbach resonances in ultracold Na~+~NaLi  \cite{Son:22,Park:23b,Karman:23} and K~+~NaK \cite{Yang:19,Zhen:22}  mixtures.

\subsection{Collisions of  $^2\Sigma$ molecules  with  $^1\mathrm{S}_0$ atoms}

Before introducing the TRAM basis, we will briefly  review the key aspects of quantum scattering theory as they apply to ultracold atom-molecule collisions in the presence of an external magnetic field.  The reader is referred to Refs.~\cite{Tscherbul:18b,Krems:19}  for a detailed account of the theory. 

The Hamiltonian of a $^{2}\Sigma$ diatomic molecule colliding with a spherically symmetric  $^1\mathrm{S}_0$ atom in a magnetic field may be written as \textcolor{red}{(using atomic units, in which $\hbar=1$)} \cite{Volpi:02,Krems:04,Tscherbul:10,Koyu:22,Tscherbul:18b}
\begin{equation}\label{eq:H}
\hat{H}=-\frac{1}{2\mu R}\frac{\partial^2}{\partial R^2}R + \frac{\hat{l}^2}{2\mu R^2}
+ \hat{V}(r,R,\theta)
+ \hat{H}_\text{mol},
\end{equation}
where $\theta$ is the  angle between the Jacobi vectors $\mathbf{R}$ and $\mathbf{r}$, which span the configuration space of the  atom-molecule collision complex,   $\mu$ is the reduced mass of the triatomic complex, and $\hat{l}^2$ the squared orbital angular momentum for the collision. The atom-molecule interaction potential $\hat{V}$ is a scalar function of the atom-molecule center of mass separation $R=|\mathbf{R}|$,  the internuclear distance of the diatomic molecule $r=|\mathbf{r}|$, and $\theta$. 
We will adopt the rigid-rotor approximation by setting $r=r_e$, where $r_e$ is the equilibrium distance of the diatomic molecule.

Here, we focus on the simplest yet common example of hyperfine structure, which arises in $^2\Sigma$ molecules bearing a single magnetic nucleus, such as laser-coolable  SrF, CaF, YbF, and YO molecules \cite{Barry:14,McCarron:18,Truppe:17,Anderegg:18,Anderegg:19,Ding:20,Wu:21,Burau:23}.
    The internal structure of such molecules and their interaction with an external magnetic field are described by the Hamiltonian, \textcolor{red}{which consists of three parts}
\begin{equation}{\label{Hmol}}
\hat{H}_\text{mol}= \hat{H}_\text{mol}^\text{rot} + \hat{H}_\text{mol}^\text{spin} + \hat{H}_\text{mol}^\text{spin-rot},
\end{equation}
where
\begin{equation}{\label{Hrot}}
\hat{H}_\text{mol}^\text{rot}  = B_e\hat{N}^2
\end{equation}
 is the rotational part,  $\hat{\mathbf{N}}$ is the rotational angular momentum of the molecule and $B_e$ is the rotational constant. The spin part, which only depends on the electron and nuclear spin operators, is given by
\begin{equation}{\label{Hmol_spin}}
 \hat{H}_\text{mol}^\text{spin}  = g_S\mu_0 B \hat{S}_Z +  (b+c/3)\hat{\mathbf{I}}\cdot\hat{\mathbf{S}},
\end{equation}
where the first term on the right-hand side is the Zeeman Hamiltonian, $\mu_0$ is the Bohr magneton,  $g_S\simeq 2.0$ is the electron spin $g$-factor, and $B$ is the magnitude of the external magnetic field,  which defines the quantization axis of the space-fixed (SF) coordinate frame. 

In Eq.~\eqref{Hmol_spin} $\hat{\mathbf{S}}$ and $\hat{\mathbf{I}}$ are the electron and nuclear spin \textcolor{red}{operators (the eigenvalues of $\hat{S}^2$ are given by $S(S+1)$ with $S=1/2$} for $^2\Sigma$ molecules).
The isotropic (or Fermi contact) hyperfine interaction  \cite{Brown:03} is given by the last term in Eq.~\eqref{Hmol_spin}, where $a=b+c/3$ is the corresponding hyperfine constant expressed via the constants $b$ and $c$ introduced by Frosch and Foley  \cite{Brown:03}.

  The interaction Hamiltonian couples the rotational and spin degrees of freedom
\begin{multline}{\label{Hmol_spin_rot}}
 \hat{H}_\text{mol}^\text{spin-rot} =  \gamma_\text{sr} \hat{\mathbf{N}}\cdot \hat{\mathbf{S}} + \frac{c\sqrt{6}}{3}\biggl{(}\frac{4\pi}{5}\biggr{)}^{1/2}
 \\ \times \sum_{q=-2}^2 (-1)^qY_{2-q}(\theta_r,\phi_r)
          [\hat{\mathbf{I}}\otimes \hat{\mathbf{S}}]^{(2)}_q.
\end{multline}
where   the electron spin-rotation interaction $\gamma_\text{sr}\hat{\mathbf{N}}\cdot \hat{\mathbf{S}}$ is parametrized by the coupling constant $\gamma_\text{sr}$, and the anisotropic hyperfine interaction by the constant $c$. 
 The spherical harmonics $Y_{2-q}$ depend on the angles $\theta_r$ and $\phi_r$, which specify the orientation of the molecular axis in the SF frame.
 We neglect the  nuclear spin-rotational interaction $C\hat{\mathbf{I}}\cdot\hat{\mathbf{N}}$, which for YbF is three orders of magnitude smaller than the electron spin-rotation interaction \cite{Tscherbul:07}.

\begin{figure}[t!]
	\centering
	\includegraphics[width=0.85\columnwidth, trim = 0 0 0 0]{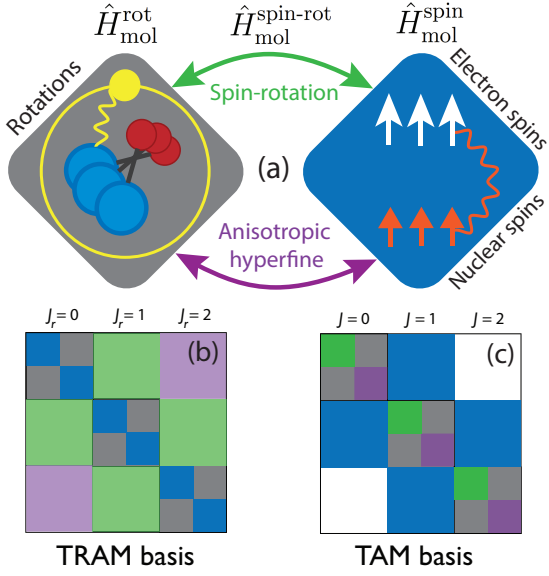}
	\caption{(a) Schematic representation of the total molecular Hilbert space as a direct product of two subspaces corresponding to  molecular rotations (left diamond) and to the electron and nuclear spin degrees of freedom (right diamond). 
	\textcolor{red}{The yellow and
orange wavy lines represent, respectively, the anisotropy of the
electrostatic atom-molecule interaction potential (left diamond) and
the isotropic (Fermi contact) hyperfine interaction (right diamond).}
	Matrix representation of molecular Hamiltonian in the TRAM basis (b) and in the TAM basis (c). The color coding scheme is as follows: grey (rotational kinetic energy and adiabatic interaction potential), blue (isotropic hyperfine and Zeeman interactions), green and violet (electron spin-rotation interaction and anisotropic hyperfine interactions).}
	\label{fig:cartoon}
\end{figure}

As illustrated in Fig.~\ref{fig:cartoon}, the commuting operators $\hat{H}_\text{mol}^\text{rot} $ and  $\hat{H}_\text{mol}^\text{spin}$ act in different subspaces of the total Hilbert space of the molecule, which is a direct product of rotational and spin  subspaces (with the latter including both the electron and nuclear spin subspaces).

Our goal is to solve the time-independent Schr\"odinger equation $\hat{H}\ket{\Psi}=E\ket{\Psi}$ with the Hamiltonian given by Eq.~(\ref{eq:H}).  
To this end we expand the solution at fixed total energy $E$ in channel functions $\ket{\Phi_n}$ 
\begin{equation}
\label{eq:psi}
|\Psi\rangle = \frac{1}{R} \sum^{}_{n}F_{n}(R) \ket{\Phi_n}, 
\end{equation}
where $F_n(R)$ are the radial functions, which satisfy  the standard coupled-channel (CC) equations \textcolor{red}{(in  atomic units, $\hbar=1$)}
\begin{multline}\label{CCeqs}
 \Big[ \frac{d^2}{dR^2}+2\mu E \Big] F_{n}(R) \\ =2\mu
 \sum_{n'} \langle \Phi_n | \hat{V} (R,\theta)+ \frac{\hat{{l}}^2}{2\mu R^2} + \hat{H}_\text{mol}   | \Phi_{n'}\rangle  
  F_{n'}(R)
\end{multline}
The asymptotic behavior of the radial functions defines the scattering $S$-matrix, from which all scattering observables can be obtained, including differential and integral cross sections, transition rates, and collision lifetimes.

\subsubsection{Matrix elements  the  TRAM basis: Intramolecular rotational, hyperfine, and Zeeman interactions}

 We expand the solution of the time-independent Schr\"odinger equation (\ref{eq:psi}) in a space-fixed TRAM basis
\begin{equation}
\label{TRAMbasis}
 \ket{\Phi_n} =   |(Nl)J_rM_r\rangle |n_s\rangle,
\end{equation}
where  $ |(Nl)J_rM_r\rangle$ are  the eigenstates  of  the {\it  total rotational  angular momentum} (TRAM) of the collision complex  $\hat{\mathbf{J}}_r= \hat{\mathbf{N}}  + \hat{\mathbf{l}} $, and its projection on the $Z$ axis $\hat{J}_{r_Z}$, and $ |n_s\rangle$  are the basis functions for the electron and nuclear spin degrees of freedom (see below). 

 The eigenstates of ${\hat{J}_r}^2$ and $\hat{J}_{r_Z}$ are obtained by coupling the eigenstates of  $\hat{N}^2$ and $\hat{N}_Z$ and those of  $\hat{l}^2$ and $\hat{l}_Z$ \cite{Zare:88}
\begin{equation}
\label{eq:totjbasis}
 |(Nl)J_rM_r\rangle = \sum_{M_N,m_l}  \langle NM_N l m_l | J_r M_r\rangle
| NM_N\rangle |l m_l\rangle,
\end{equation}
where $ |l m_l\rangle$ are the eigenstates of $\hat{l}^2$ and $\hat{l}_Z$, $\ket{NM_N}$ are those of $|\hat{\mathbf{N}}|^2$ and $\hat{N}_Z$, and $\langle j_1m_1j_2 m_2|jm\rangle$ are the Clebsch-Gordan coefficients.  Note that  $\hat{\mathbf{J}}_r= \hat{\mathbf{N}}  + \hat{\mathbf{l}}$  is different from the  total angular momentum operator  $\mathbf{\hat{J}} = \hat{\mathbf{N}} + \hat{\mathbf{l}} + \hat{\mathbf{S}} + \hat{\mathbf{I}} = \mathbf{\hat{J}}_r + \mathbf{\hat{S}} + \mathbf{\hat{I}}$ unless $I=S=0$.


Below we will evaluate the matrix elements of the scattering Hamiltonian in the  TRAM basis. We also discuss the structure of the coupling matrix elements, which makes clear  the unique advantages offered by this basis for incorporating hyperfine structure and magnetic fields in quantum scattering calculations on molecular collisions in external fields.

 We begin with the  Hamiltonian of the diatomic molecule \eqref{Hmol}.
 Taking advantage of the structure of the TRAM basis, which is a direct product of basis functions for  the rotational and spin degrees of freedom, we find
\begin{multline}\label{mel_Hmol}
\langle (Nl)J_rM_r | \langle n_s| \hat{H}_\text{mol} | (N'l')J_r'M_r' \rangle | n_s'\rangle
 \\ 
 =\delta_{n_s n_s'} \langle (Nl)J_rM_r| \hat{H}_\text{mol}^\text{rot}  | (N'l')J_r'M_r' \rangle  
 \\ 
+ \delta_{NN'}\delta_{ll'}\delta_{J_rJ_r'}\delta_{M_r M_r'} \langle n_s| \hat{H}_\text{mol}^\text{spin}   | n_s'\rangle   \\ 
+ \langle (Nl)J_rM_r | \langle n_s| \hat{H}_\text{mol}^\text{spin-rot}  | (N'l')J_r'M_r' \rangle | n_s'\rangle.
\end{multline}
We observe that the purely rotational part of the Hamiltonian $\hat{H}_\text{mol}^\text{rot} $ is diagonal in the spin degrees of freedom, whereas the spin part $ \hat{H}_\text{mol}^\text{spin} $  (including the Zeeman interaction) is diagonal in  the rotational degrees of freedom, as expected. The spin-rotation interaction $\hat{H}_\text{mol}^\text{spin-rot} $  couples the rotational and spin degrees of freedom. 

To proceed, we need to specify the basis ket vectors $\ket{n_s}$ for the spin degrees of freedom. Here,  as in our previous work \cite{Tscherbul:07}, we use the fully uncoupled spin basis set $\ket{n_s}=\ket{SM_S}\ket{IM_I}$, although the coupled hyperfine basis  $\ket{n_s}=|Fm_F\rangle$ could be used  as well.
The first term on the right-hand side of Eq.~\eqref{mel_Hmol} is diagonal since the basis states $|(Nl)J_r M_r \rangle $ are eigenstates of $\hat{N}^2$
\begin{multline}\label{mel_Hmol_term1}
\langle (Nl)J_rM_r | \langle SM_S | \langle IM_I | \hat{H}_\text{mol}^\text{rot} | (N'l')J_r'M_r' \rangle | SM_S'\rangle |IM_I'\rangle
\\ 
=\delta_{M_I M_I'} \delta_{M_S M_S'} \delta_{NN'}\delta_{ll'}  \delta_{J_rJ_r'}\delta_{M_r M_r'}  B_e N(N+1).
\end{multline}
The matrix element of $\hat{H}_\text{mol}^\text{spin} = a \mathbf{\hat{I}}\cdot\mathbf{\hat{S}} +g_S\mu_0  B\hat{S}_Z$ in the second term is straightforward to derive since our spin basis states $|SM_S\rangle|IM_I\rangle$ are eigenstates of $\hat{S}_Z$ and $\hat{I}_Z$, and thus (see, e.g., Eq.~(7) of Ref.~\cite{Tscherbul:07})
  \begin{multline}\label{mel_Hmol_term2}
\langle (Nl)J_rM_r | \langle SM_S | \langle IM_I |  \hat{H}_\text{mol}^\text{spin}  | (N'l')J_r'M_r' \rangle | SM_S'\rangle |IM_I'\rangle
\\ =  \delta_{NN'}\delta_{ll'}  \delta_{J_rJ_r'}\delta_{M_r M_r'}  
a \bigl{[}  \delta_{M_IM_I'}\delta_{M_SM_S'}M_IM_S \\ +  \frac{1}{2}\delta_{M_I,M_I'\pm 1}\delta_{M_S,M_S'\mp 1} C_\pm (I,M_I')C_\mp (S,M_S') \bigr{]},
\end{multline}
 where $C_\pm(J,M) = [J(J+1)-M(M \pm 1)]^{1/2}$. Note that  the spin Hamiltonian $\hat{H}_\text{mol}^\text{spin}$ is diagonal in $J_r$ because the isotropic hyperfine and Zeeman interactions do not act on the rotational degrees of freedom.  This is a key difference between the TRAM basis and the more familiar  total angular momentum basis, in which the Zeeman interaction  couples basis states with different $J$ \cite{Tscherbul:10,Koyu:22}.

 
Finally, the  spin-rotation coupling matrix element [the third term in Eq.~\eqref{mel_Hmol}]  is the sum of the matrix elements of the electron spin-rotation interaction $\hat{H}_\text{mol}^\text{esr}= \gamma_\text{sr}\mathbf{\hat{N}}\cdot \mathbf{\hat{S}} $ and of the anisotropic hyperfine interaction  $\hat{H}_\text{mol}^\text{ahf}=\frac{c\sqrt{6}}{3}{(}\frac{4\pi}{5}{)}^{1/2}\sum_{q=-2}^2 (-1)^qY_{2-q}(\theta_r,\phi_r)
          [\hat{\mathbf{I}}\otimes \hat{\mathbf{S}}]^{(2)}_q $ [see Eq.~\eqref{Hmol_spin_rot}].

 The matrix elements of the electron spin-rotation interaction in the TRAM basis  are given by (see Appendix A)
 \begin{multline}\label{mel_Hmol_term3sr}
\langle (Nl)J_rM_r | \langle S M_S | \langle IM_I | \hat{H}_\text{mol}^\text{esr} | (N'l')J_r'M_r' \rangle | SM_S'\rangle |IM_I'\rangle 
\\
= \delta_{M_I M_I'}\delta_{ll'}\delta_{NN'} \gamma_\text{sr} p_3(N)p_3(S) 
 [(2J_r+1)(2J_r'+1)]^{1/2}
 \\  \times
  \sum_p (-1)^{p} (-1)^{J_r-M_r}(-1)^{N+l+J_r'+1} (-1)^{S-M_S} \sixj{N}{J_r'}{J_r}{N}{l}{1}
 \\ \times
 \threejm{J_r}{-M_r}{1}{p}{J_r'}{M_r'}
 \threejm{S}{-M_S}{1}{-p}{S}{M_S'} 
\end{multline}
where $p_3(X)=[(2X+1)X(X+1)]^{1/2}$.  It follows from Eq.~\eqref{mel_Hmol_term3sr} that the electron spin-rotation interaction couples TRAM basis states with $J_r - J_r'=\pm 1$. This interaction conserves the value of $M_r+M_S$ as well as the total angular momentum projection $M=M_r+M_S+M_I$.  This coupling is typically much weaker than either the rotational energy scale or the Zeeman interaction at moderate and high magnetic fields.
Nevertheless, it plays an important role in collisions of  $^2\Sigma$ molecules with structureless atoms \cite{Krems:03,Krems:04}. In order to account for this interaction,  it is therefore necessary to include at least two  $J_r$ blocks ($J_r=0{-}1$ for the initial $N=0$ states). 

The matrix elements of the intramolecular anisotropic hyperfine interaction take the form (see Appendix B)
\begin{widetext}
 \begin{multline}\label{mel_Hmol_term3ahf}
\langle (Nl)J_rM_r | \langle S M_S | \langle IM_I | \hat{H}_\text{mol}^\text{ahf} | (N'l')J_r'M_r' \rangle | SM_S'\rangle |IM_I'\rangle 
=  \delta_{ll'} c\frac{\sqrt{30}}{3} (-1)^{J_r-M_r+l+J_r'}
 p_3(I) p_3(S)  [(2J_r+1)(2J_r'+1)]^{1/2}  \\  \times
 [(2N+1)(2N'+1)]^{1/2} \sixj{N}{J_r'}{J_r}{N'}{l}{2}  \threejm{1}{M_I-M_I'}{1}{M_S-M_S'}{2}{M_r-M_r'}  \\ \times
 \threejm{J_r}{-M_r}{2}{M_r-M_r'}{J_r'}{M_r'}
 \threejm{N}{0}{2}{0}{N'}{0} 
  \threejm{I}{-M_I}{1}{M_I-M_I'}{I}{M_I'}
 \threejm{S}{-M_S}{1}{M_S-M_S'}{S}{M_S'}
\end{multline}
\end{widetext}
The anisotropic hyperfine interaction couples  the states with $J_r - J_r'=\pm2$, and with $N-N'=\pm 2$ but conserves the total angular momentum projection $M$.  Thus,  in order to account for this interaction, it is necessary to include at least three lowest $J_r$ blocks ($J_r=0{-}2$ for the initial $N=0$ states). 

The spectroscopic constants of YbF($^2\Sigma$)  are (in units of cm$^{-1}$): $c=2.84875\times 10^{-3}$, $\gamma_\text{sr}=4.4778\times 10^{-4}$,  $b=4.72983\times 10^{-3}$, and  $B_e = 0.24129$. The anisotropic hyperfine interaction is thus 6.4 times stronger than the electron spin-rotation interaction, but two times weaker  than the isotropic (Fermi contact) hyperfine interaction  parametrized by $a=b+c/3=5.6794\times 10^{-3}$ cm$^{-1}$.

\subsubsection{Matrix elements  the  TRAM basis: Orbital angular momentum and  interaction potential}

To complete the parametrization of CC equations \eqref{CCeqs} in the TRAM basis, we need to evaluate  the matrix elements of the centrifugal kinetic energy and of the atom-molecule  interaction potential.
As the TRAM basis functions \eqref{TRAMbasis} are eigenfunctions of $\hat{l}^2$, the centrifugal kinetic energy has only diagonal matrix elements
 \begin{multline}\label{mel_Hmol_centrifugalKE}
\langle (Nl)J_rM_r | \langle n_s |   \frac{\hat{{l}}^2}{2\mu R^2}  | (N'l')J_r'M_r' \rangle | n_s'\rangle 
= \delta_{NN'}\delta_{ll'}  \delta_{J_rJ_r'}
\\ \times
\delta_{M_r M_r'}  \delta_{n_s n_s'} \frac{l(l+1)}{2\mu R^2}.
\end{multline}

The adiabatic interaction potential between a $^2\Sigma$ molecule and a $^1S$ atom is independent of the electron and nuclear spins, so its matrix elements are diagonal in $n_s$. Expanding the potential in Legendre polynomials as $V(R,\theta)=\sum_\lambda V_\lambda(R)P_\lambda(\cos\theta)$ and using the Wigner-Eckart theorem, we find the matrix elements  \cite{Lester:76}
\begin{multline}\label{InteractionPotentialMatrix}
\langle  (Nl)J_rM_r | \langle n_s | V(R,\theta) | (N'l')J_r'M_r'\rangle | n_s'\rangle
=\delta_{n_s n_s'} \delta_{J_rJ_r’}\\ 
\times \delta_{M_rM_r’} (-1)^{J_r+N+N'}
 [(2N+1)(2N’+1)(2l+1)(2l’+1)]^{1/2}
 \\ \times
\sum_\lambda V_\lambda(R) 
\begin{Bmatrix}
N&l&J\\
l’& N’ &\lambda
\end{Bmatrix}
\begin{pmatrix}
N&\lambda&N’\\
0& 0 &0
\end{pmatrix}
\begin{pmatrix}
l&\lambda&l’\\
0& 0 &0
\end{pmatrix}.
\end{multline}
The matrix elements are diagonal in $J_r$. Because the couplings between the different $J_r$ blocks due to the spin-rotation interactions  are weak (see the previous section), the scattering Hamiltonian is approximately block-diagonal in the TRAM representation as shown in Fig.~\ref{fig:cartoon}(b). Thus, one might expect that  numerical solutions of CC equations may be efficiently obtained with only the few lowest $J_r$ blocks retained in the basis. As shown in Sec.~III below, this expectation turns out to be true. This advantage of the TRAM basis is similar to that provided by the total angular momentum representation \cite{Tscherbul:10,Suleimanov:12}.  


The matrix elements \eqref{InteractionPotentialMatrix}  have a simple form, which does not increase in complexity as additional hyperfine or electron spin degrees of freedom are added. The underlying reason for this simplicity is that in the TRAM representation, one only couples the {\it rotational} angular momenta, on which the interaction potential actually depends. 
By contrast, in the space-fixed TAM representation, one needs to couple {\it all} angular momenta regardless of whether or not they are coupled by the interaction potential, leading to a nested hierarchy of partially coupled basis functions, whose complexity increases rapidly as more angular momenta are added \cite{Koyu:22,Suleimanov:12}.

\subsection{Collisions of  $^2\Sigma$ molecules with  $^2\mathrm{S}$ atoms}

The Hamiltonian of the collision complex formed by a $^{2}\Sigma$ molecule and an  $^2\mathrm{S}_0$ atom may be  written as
\begin{equation}\label{eq:H2}
\hat{H}=-\frac{1}{2\mu R}\frac{\partial^2}{\partial R^2}R + \frac{\hat{l}^2}{2\mu R^2}
+ \hat{V}_\text{int} 
+ \hat{H}_\text{mol} + \hat{H}_\text{atom} + \hat{V}_\text{mdd},
\end{equation}
where all the terms except for the interaction potential operator $\hat{V}_\text{int}$ and the new terms $\hat{H}_\text{atom}$ and $\hat{V}_\text{mdd}$ (see below) have the same meaning as in Eq.~\eqref{eq:H}. 
The Hamiltonian (\ref{eq:H2})  differs from Eq.~\eqref{eq:H} in three significant respects.

First, the $^2\mathrm{S}_{1/2}$ atomic collision partner (such as an alkali-metal atom) has internal structure described by the hyperfine-Zeeman Hamiltonian
\begin{equation}{\label{Hatom}}
 \hat{H}_\text{atom}  = g_S\mu_0 B \hat{S}_{a_Z} + A_a\hat{\mathbf{I}}_a\cdot\hat{\mathbf{S}}_a,
\end{equation}
where $\hat{\mathbf{S}}_a$ and $\hat{\mathbf{I}}_a$ are the atomic electron and nuclear spin operators, and $A_a$ is the atomic hyperfine constant.
For simplicity, we will neglect the dependence of atomic and molecular hyperfine constants on $R$, $r$, and $\theta$ as well as tensor hyperfine couplings of the form 
\textcolor{red}{$\hat{\mathbf{S}}\cdot \mathsf{T}(R,\theta,r)\cdot  \hat{\mathbf{I}}_a$, where $\mathsf{T}(R,\theta,r)$ is a second-rank  tensor that describes the coupling of the molecule's electron spin with the nuclear spin of the atom. Other forms of the tensor hyperfine coupling are possible, such as $\hat{\mathbf{S}}_a\cdot \mathsf{T}_a(R,\theta,r)\cdot  \hat{\mathbf{I}}$, which describes the coupling of the atom's electron spin with the nuclear spin of the diatomic molecule.
  These expressions can be expanded in either Cartesian or spherical tensor products (see Refs.~\cite{SchweigerBook,Upadhyay:19} for more details).}
While these tensor  interactions are typically weaker than those already included in Eqs.~\eqref{Hmol} and (\ref{Hatom}) (see, e.g., \cite{Tscherbul:09d,Tscherbul:11c}),  they can become substantial in the short-range collision complex region \cite{Jachymski:22}. These  interactions can be handled by evaluating their matrix elements in the TRAM basis, and will result in additional mixing between the states of different $J_r$ similar in form to the couplings induced by the magnetic dipole-dipole interaction as described below.

Second, the  interaction potential between a $^2\Sigma$ molecule and a $^2\mathrm{S}$ atom  is no longer a single scalar function of the internal coordinates $R$, $r$, and $\theta$ as was the case for interactions with structureless atoms (see Sec. IIA), but depends on the eigenvalues \textcolor{red}{$S_T(S_T+1)$ of $\hat{S}_T^2$, where  $\hat{\mathbf{S}}_T = \hat{\mathbf{S}}+\hat{\mathbf{S}}_a$ is the total spin of the atom-molecule collision complex}  \cite{Morita:18}
\begin{equation}{\label{V_Sdependent}}
\hat{V}_\text{int}  = \sum_{S_T, M_{S_T}} V^{S_T}(R,r,\theta) |S_T M_{S_T}\rangle \langle S_T M_{S_T}|
\end{equation}
where $V_{S_{T}}(R,r,\theta) $ are the adiabatic potential energy surfaces (PESs) for the singlet $(S_T=0)$ and triplet $(S_T=1)$ electronic states.  These PESs are typically very different at short range, where strong exchange interactions lead to large singlet-triplet energy gaps, as in the case of two interacting hydrogen or alkali-metal atoms \cite{Karplus:70,Chin:10}.

Finally, the magnetic dipole-dipole interaction between the electron spins of the open-shell collision partners takes the form \cite{Krems:04,Janssen:11,Suleimanov:12,Morita:18}
\begin{equation}\label{Vmdd}
\hat{V}_\text{mdd}  =  -\left( \frac{24\pi}{5}\right)^{1/2} \frac{\alpha^2}{R^3} \sum_{q} (-1)^q Y_{2,-q}(\hat{R})[\hat{\mathbf{S}}\otimes\hat{\mathbf{S}}_a]^{(2)}_q  
\end{equation}
where the spherical harmonic $Y_{2,-q}(\hat{R})$ depends on the orientation of vector $\hat{R}=\mathbf{R}/R$ in the space-fixed frame, and  $[\hat{\mathbf{S}}\otimes\hat{\mathbf{S}}_a]^{(2)}_q $ is a second-rank tensor product of the molecular and atomic electron spin operators.

The TRAM basis for collisions of  $^2\Sigma$ molecules with  $^2\mathrm{S}$ atoms is a direct product of basis states for the structureless atom collision problem [Eq.~\eqref{TRAMbasis}] with atomic spin basis functions $|n_s^{(a)}\rangle$
\begin{equation}
\label{TRAMbasis2}
 \ket{\Phi_n} =   |(Nl)J_rM_r\rangle |n_s\rangle |n_s^{(a)}\rangle.
\end{equation}
As before, we choose the molecular  and atomic spin basis functions in the uncoupled angular momentum representation \cite{Krems:04,Volpi:02}
\begin{align}\label{spin_fun}\notag
|n_s\rangle &= |S M_{S}\rangle | I M_{I}\rangle = |S M_{S}  I M_{I}\rangle \\
|n_s^{(a)}\rangle &= |S_a M_{S_a}\rangle | I_a M_{I_a}\rangle = |S_a M_{S_a}  I_a M_{I_a}\rangle. 
\end{align}
Alternatively, one could use a coupled representation, where $|n_s\rangle=|(IS)F M_{F}\rangle$ and  $|n_s^{(a)}\rangle=|(I_aS_a)F_a M_{F_a}\rangle$.

 The  matrix elements of the molecular Hamiltonian (\ref{Hmol}) and of the centrifugal kinetic energy are  diagonal in atomic spin quantum numbers $n_s^{(a)}$. The  expressions  for these matrix elements are identical to those already presented in Sec.~IIA.
We now proceed to evaluate the matrix elements of  the three remaining terms \eqref{Hatom}--\eqref{Vmdd}.

The matrix elements of the atomic Hamiltonian are diagonal in all molecular quantum numbers
\begin{multline}{\label{mel_Hatom}}
\langle  (Nl)J_rM_r  \langle n_s | \langle n_s^{(a)} |  \hat{H}_\text{atom} |(N'l')J_r' M_r' \rangle |n_s' \rangle |n_s^{(a)\prime}\rangle  
\\
 = \delta_{NN'}\delta_{ll'}\delta_{J_rJ_r'}\delta_{M_rM_r'}\delta_{n_sn_s'}\langle n_s^{(a)} |  \hat{H}_\text{atom} | n_s^{(a)\prime}
\rangle 
\end{multline}
Choosing the fully uncoupled representation for the atomic basis functions, the right-hand side evaluates to, in close analogy with Eq.~\eqref{mel_Hmol_term2}
  \begin{multline}\label{mel_Hatom2}
 \langle S_aM_{S_a} | \langle I_a M_{I_a} |  \hat{H}_\text{atom}  |  S_a M_{S_a}'\rangle |I_a M_{I_a}'\rangle
=    \delta_{M_{I_a}M_{I_a}'}\delta_{M_{S_a} M_{S_a}'} \\ \times g_S\mu_0 B M_{S_a} 
+ A_a \bigl{[} \delta_{M_{I_a}M_{I_a}'} \delta_{M_{S_a} M_{S_a}'} M_IM_S 
\\ +  \frac{1}{2}\delta_{M_{I_a},M_{I_a}'\pm 1}\delta_{M_{S_a},M_{S_a}'\mp 1} C_\pm (I_a,M_{I_a}')C_\mp (S_a,M_{S_a}') \bigr{]}.
\end{multline}

The interaction potential between a $^2\Sigma$ molecule and a $^2$S atom  is expressed in terms of projectors on  eigenstates  of the total electron spin $|S_T M_T\rangle$ of the atom-molecule system  \eqref{V_Sdependent}. Its matrix elements  in the TRAM basis factorize into the rotational and electron spin parts
\begin{widetext}
\begin{multline}{\label{mel_V_Sdependent}}
\langle (Nl)J_rM_r | \langle  S M_{S} I M_{I} | \langle S_a M_{S_a}  I_a M_{I_a} | \hat{V}_\text{int} |(N'l')J_r' M_r' \rangle |S M_{S}' I M_{I}'  \rangle  S_a M_{S_a}'  I_a M_{I_a}' \rangle  = \delta_{M_I M_I'} \delta_{M_{I_a} M_{I_a}'}
\\ \times
 \sum_{S_T, M_{S_T}} 
 \langle (Nl)J_rM_r | \langle n_s |V^{S_T}(R,r,\theta) |(N'l')J_r' M_r' \rangle 
 \langle SM_S| \langle S_a M_{S_a} |S_T M_{S_T}\rangle \langle S_T M_{S_T}| S'M_S'\rangle  | S_a M_{S_a}'\rangle 
\end{multline}
\end{widetext}
The  total spin eigenfunctions are expressed in terms of  the uncoupled spin  functions as $|S_T M_{S_T}\rangle=\sum_{M_{S}}\langle SM_S, S_a M_{S_a}|SM_{S_T}\rangle |SM_S\rangle|S_aM_{S_a}\rangle$ \cite{Krems:04}. Multiplying the Hermitian conjugate of this expression by $| S M_{S}' \rangle |S_a'  M_{S_a}'\rangle$ and integrating over the spin degrees of freedom, the spin overlaps in Eq.~\eqref{mel_V_Sdependent} can be expressed in terms of the Clebsch-Gordan coefficients, e.g., $\langle S_T M_{S_T}| S'M_S'\rangle  | S_a M_{S_a}'\rangle =\langle S M_S',S_a M_{S_a}'|S_T M_{S_T}\rangle$. Substituting the rotational matrix elements of $V^S(R,\theta,r)$ from Eq.~(\ref{InteractionPotentialMatrix}), we obtain the final result
\begin{widetext}
\begin{multline}\label{mel_V_Sdependent_final}
\langle (Nl)J_rM_r | \langle  S M_{S} I M_{I} | \langle S_a M_{S_a}  I_a M_{I_a} | \hat{V}_\text{int} |(N'l')J_r' M_r' \rangle |S M_{S}' I M_{I}'  \rangle | S_a M_{S_a}'  I_a M_{I_a}' \rangle  = \delta_{M_I M_I'} \delta_{M_{I_a} M_{I_a}'}
\\ \times
 \delta_{J_rJ_r’}\delta_{M_rM_r’} (-1)^{J_r+N+N'}
 [(2N+1)(2N’+1)(2l+1)(2l’+1)]^{1/2} \sum_{S_{T},M_{S_T}}(-1)^{2(S-S_a+M_{S_T})}(2S_T+1) 
 \\ \times
 \threejm{S}{M_S}{S_a}{M_{S_a}}{S_T}{-M_{S_T}}
  \threejm{S}{M_S'}{S_a}{M_{S_a}'}{S_T}{-M_{S_T}}
\sum_\lambda V^{S_T}_\lambda(R,r) 
\begin{Bmatrix}
N&l&J\\
l’& N’ &\lambda
\end{Bmatrix}
\begin{pmatrix}
N&\lambda&N’\\
0& 0 &0
\end{pmatrix}
\begin{pmatrix}
l&\lambda&l’\\
0& 0 &0
\end{pmatrix},
\end{multline}
\end{widetext}
where the spin-dependent Legendre expansion coefficients are defined by  $V^{S_T}(R,\theta,r)=\sum_\lambda V^{S_T}_\lambda(R,r) P_\lambda(\cos\theta)$ for each $S_T=|S-S_a|,\ldots,S+S_a$.
The matrix of the  electrostatic interaction potential is diagonal in $J_r$, which is significant because the anisotropic atom-molecule interactions can be block-diagonalized, enabling the scattering problem to be solved independently for each $J_r$ even in the presence of the Zeeman and isotropic hyperfine interactions (since, as shown above, these  interactions are diagonal in $J_r$).  The only interactions that couple the states of different $J_r$ are the spin-rotation interactions (see above and Fig.~1), which are weak, so only a few values of $J_r$ are sufficient to achieve numerical convergence of scattering observables, as demonstrated below.


The interaction potential matrix is also diagonal in the atomic and molecular nuclear spin projections, since the potential does not depend on the nuclear spin operators. The interaction potential couples the states with different  $M_S$ and $M_{S_a}$ and the same $M_S=M_{S} + M_{S_a}$ due to the difference between the potentials with different  $S_T$ at short range (this is analogous to the spin-exchange interaction in alkali-metal dimers). 
The only exception are the states with  the maximum possible $M_S$ and $M_{S_a}$, which correspond to the fully stretched basis states $|S,M_S=S\rangle$ and $|S_a,M_{S_a}=S\rangle$ with  $S_T=S+S_a$.  These states  occur in collisions of fully spin-polarized molecules and/or atoms, whose collision dynamics can often be adequately described by a single, high-spin PES  \cite{Suleimanov:12,Morita:18}.

Finally, the magnetic dipolar interaction is a contraction of tensor operators \eqref{Vmdd}, which depend on the orientation  of the atom-molecule axis in the SF frame [via the term $Y_{2,-q}(\hat{R})$] and on the spin degrees of freedom [via the term $[\hat{\mathbf{S}}\otimes\hat{\mathbf{S}}_a]^{(2)}_q$]. The matrix element of Eq.~\eqref{Vmdd}  in the TRAM basis  then takes the form
\begin{widetext}
\begin{multline}\label{Vmdd_mel}
\langle (Nl)J_rM_r | \langle  S M_{S} I M_{I} | \langle S_a M_{S_a}  I_a M_{I_a} | \hat{V}_\text{mdd} |(N'l')J_r' M_r' \rangle |S M_{S}' I M_{I}'  \rangle | S_a M_{S_a}'  I_a M_{I_a}' \rangle  = 
    -\left( \frac{24\pi}{5}\right)^{1/2}  \frac{\alpha^2}{R^3}  \delta_{M_I M_I'} \delta_{M_{I_a} M_{I_a}'} \\ \times \sum_{q} (-1)^q
    \langle (Nl)J_rM_r  |  Y_{2,-q}(\hat{R})|(N'l')J_r' M_r' \rangle  
    \langle  S M_{S} | \langle S_a M_{S_a} |  [\hat{\mathbf{S}}\otimes\hat{\mathbf{S}}_a]^{(2)}_q | S M_{S}' \rangle  | S_a M_{S_a}'  \rangle
\end{multline}
\end{widetext}

Using  the  definition of the tensor product  \cite{Zare:88} to evaluate the  matrix element of $[\hat{\mathbf{S}}\otimes\hat{\mathbf{S}}_a]^{(2)}_q $ and then applying the Wigner-Eckart theorem, one obtains \textcolor{red}{(see Appendix~C)}
\begin{widetext}
\begin{multline}\label{Vmdd_mel2}
\langle (Nl)J_r M_r | \langle  S M_{S} I M_{I} | \langle S_a M_{S_a}  I_a M_{I_a} | \hat{V}_\text{mdd} |(N'l')J_r' M_r' \rangle |S M_{S}' I M_{I}'  \rangle | S_a M_{S_a}'  I_a M_{I_a}' \rangle
   =   \frac{-\sqrt{30}\alpha^2}{R^3}  \delta_{M_I M_I'} \delta_{M_{I_a} M_{I_a}'} \delta_{NN'}
      \\ \times (-1)^{2J_r-M_r+N + l' +l}[(2J_r+1)(2J_r'+1)]^{1/2}
      \threejm{J_r}{-M_r}{2}{M_r-M_r'}{J_r'}{M_r'} 
\sixj{l}{J_r'}{J_r}{l'}{N}{2}
   [(2l+1)(2l'+1)]^{1/2} \threejm{l}{0}{2}{0}{l'}{0} p_3(S) p_3(S_a)
    \\ \times
    \threejm{1}{M_S-M_S'}{1}{M_{S_a}-M_{S_a}'}{2}{M_r-M_r'}
       (-1)^{S-M_S+S_a-M_{S_a}} \threejm{S}{-M_S}{1}{M_{S}-M_{S}'}{S}{M_S'}  \threejm{S_a}{-M_{S_a}}{1}{M_{S_a}-M_{S_a}'}{S_a}{M_{S_a}'}
    \end{multline}
    \end{widetext}
The magnetic dipolar interaction is seen to be diagonal in the nuclear spin quantum numbers and to couple the TRAM basis states with values of $J_r$ differing by two. The minimum TRAM basis for the magnetic dipolar interaction should  therefore  contain at least three $J_r$ blocks ($J_r=0-2$ for the initial $N=0$ states) as in the case of the anisotropic hyperfine interaction considered above.  Unlike the atom-molecule interaction potential, the magnetic dipolar interaction does not separately conserve the rotational and spin angular momentum projections $M_r$, $M_S$, and $M_{S_a}$, but does conserve the sum  $M_S+M_{S_a}+M_r$.

We now summarize the key advantages of the TRAM basis, which follow from the  above  discussion.

\begin{enumerate}

\item
Most of the terms in the scattering Hamiltonian, including the interaction potential \eqref{InteractionPotentialMatrix}, the interaction of the molecule and atom with an external magnetic field, and the isotropic hyperfine interaction [Eqs.~\eqref{Hmol}  and \eqref{Hatom}], are diagonal in the TRAM quantum number $J_r$. Thus, the atom-molecule scattering Hamiltonian is strongly diagonally dominant in the TRAM representation as shown in Fig.~\ref{fig:cartoon}(a).  
{While this nearly block-diagonal structure is superficially similar to the one, which occurs in the TAM basis \cite{Tscherbul:10,Koyu:22}, there are important differences. Specifically, the Zeeman interaction is the only interaction, which is  non-diagonal in $J$  \cite{Tscherbul:10,Koyu:22}.}

\item
 Couplings between the different values of $J_r$ arise {\it only}  due to the interaction of molecular rotations with the electron and/or nuclear spins.
  These interactions can originate either from within the diatomic molecule, due to the spin-rotation interaction  \eqref{Hmol_spin_rot} or from the intramolecular magnetic dipole-dipole interaction. These couplings are typically weak.

\item Because of the nearly block-diagonal  structure of the Hamiltonian in the TRAM basis, one might expect that efficient numerical solutions of CC equations may be obtained with only the few lowest $J_r$ blocks retained in the basis. As shown in Sec.~III below, this expectation turns out to be true.  This is a key advantage of the TRAM basis, which allows efficient handling of  strongly anisotropic interactions. This advantage is  similar to that provided by the total angular momentum basis \cite{Tscherbul:10,Suleimanov:12}.  

\end{enumerate}

\section{Results}

In this section, we apply the TRAM representation developed in the previous section to calculate the cross sections for ultracold collisions between YbF$(^2\Sigma)$ molecules and He atoms in the presence of an external magnetic field.  YbF   is of interest for precision searches  for the electric dipole moment of the electron \cite{Hudson:11}, and it has recently been laser cooled \cite{Lim:18}. YbF has a pronounced hyperfine structure, whose marked effect on cold and ultracold He~+~YbF collisions was   studied in our previous work using a fully uncoupled angular momentum basis \cite{Tscherbul:07}. These calculations provide a convenient benchmark, against which we will test  our TRAM approach. 

 \begin{figure}[t!]
	\centering
	\includegraphics[width=0.9\columnwidth, trim = 30 0 0 -10]{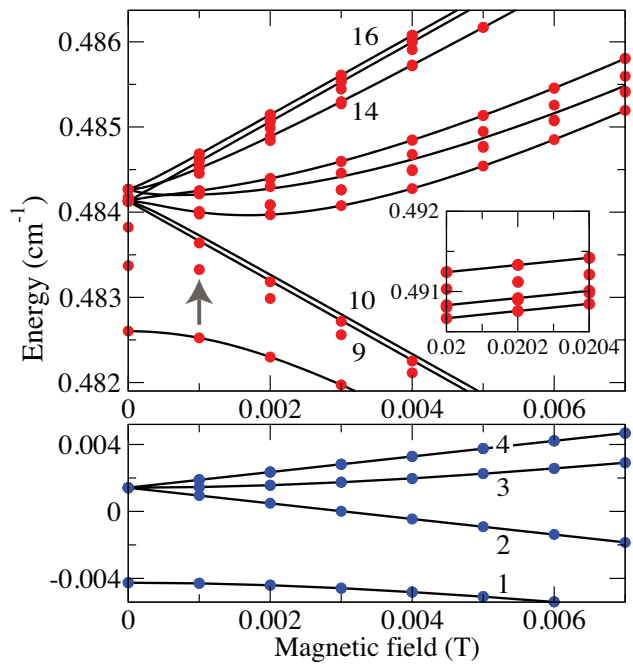}
	\caption{Hyperfine-Zeeman energy levels in the ground ($N=0$, \textcolor{red}{bottom}) and the first rotationally excited ($N=1$, \textcolor{red}{top}) manifolds of YbF$(^2\Sigma)$ plotted as a function of magnetic field. Solid lines -- physical hyperfine-Zeeman levels, circles --  eigenvalues of the molecular Hamiltonian in the TRAM basis calculated for $N^\text{max}=4$, $J_r^\text{max}=2$, and $M=1$. The inset shows a portion of the energy level spectra at higher fields. For $N=0$ the deviation of the eigenvalues of the asymptotic Hamiltonian in the TRAM basis from the physical hyperfine-Zeeman levels of YbF ($\le 0.01$\%) is much less than the size of the symbols. }
	\label{fig:energy_levels}
\end{figure}

\subsection{Hyperfine-Zeeman energy levels and unphysical states}

Figure \ref{fig:energy_levels} shows the lowest hyperfine-Zeeman energy levels of YbF($^2\Sigma$) as a function of magnetic field.
The physical energy levels shown by straight lines are obtained by diagonalization  of the  isolated YbF Hamiltonian \eqref{Hmol}  in the single-molecule basis $|NM_N\rangle |SM_S\rangle |IM_I\rangle$. The eigenvalues of the same Hamiltonian in the TRAM basis obtained by diagonalizing the matrix in Eq.~\eqref{mel_Hmol} are shown by symbols. We note that these sets of levels are not necessarily identical because the spin-rotation and anisotropic hyperfine interactions couple the states  of different $J_r$ (see Fig.~\ref{fig:cartoon}). Since our TRAM basis is truncated at a finite value of $J_r^\text{max}$, the couplings between the $J_r$-th and $(J_r+1)$-th blocks are excluded from consideration,  leading to the appearance of unphysical energy levels, just as in the case of the TAM representation \cite{Tscherbul:10,Tscherbul:12,Koyu:22}.  

In further analogy with the unphysical states that occur in the TAM representation \cite{Tscherbul:10,Tscherbul:12,Koyu:22},
the eigenvectors of the unphysical states are dominated by contributions from the highest value of $J_r$ included in the TRAM basis. Consider, e.g., the unphysical state with the largest  deviation from any physical hyperfine-Zeeman state in the $N=1$ manifold. This state is marked by an arrow in Fig.~\ref{fig:energy_levels} and its eigenvector can be expanded in the TRAM basis states $\ket{J_r,M_r,M_I,M_S}$ at $B=10^{-3}$~T
\begin{multline}\label{eigenvector}
|u\rangle =  0.83 \ket{2,2,   -\frac{1}{2}, -\frac{1}{2}} -  0.31\ket{2, 0,  \frac{1}{2}, \frac{1}{2}} 
\\+ 0.225\ket{1,1,   \frac{1}{2}, -\frac{1}{2}}     + \cdots, 
\end{multline}
where the quantum numbers that take fixed values ($N=1$ and $l=2$) have been omitted from basis kets for clarity.
We see that by far the largest contribution to the unphysical state is given by the states in the $J_r=2$ block.
This suggests that, as shown in the next Section, the unphysical states do not  affect the results of quantum scattering calculations on ultracold molecular collisions, which are determined by the  lowest $J_r$ states in the TRAM basis. 

As shown in the lower panel of Fig.~\ref{fig:energy_levels}, the $N=0$ eigenstates computed in the TRAM basis are nearly identical to the physical hyperfine-Zeeman states of YbF, with the relative deviations not exceeding  $0.01\%$. This is because the  different $J_r$ blocks are only coupled by the relatively weak, field-independent spin-rotation and anisotropic hyperfine interactions.
This is in contrast to the TAM representation, where the different $J$ blocks are coupled by the external magnetic field, causing the appearance of distinct unphysical  Zeeman states in all $N$ manifolds even in the absence of hyperfine structure \cite{Tscherbul:10,Tscherbul:12,Koyu:22}.

The $N=1$ manifold  has more  unphysical states especially at smaller magnetic fields, where the couplings between the different $J_r$ blocks are larger than the Zeeman interaction.  Even for these states, the relative deviation between the physical and unphysical energies does not exceed $0.07$\%. The largest deviation occurs for the state closest to state $\ket{9}$ marked by an arrow in Fig.~\ref{fig:energy_levels}. In addition, the density of unphysical states in the TRAM basis is significantly lower than in the TAM representation, where {\it most} of the $N=1$ states are unphysical \cite{Koyu:22}.  
The lower density of unphysical states  represents an advantage of the TRAM basis over the  TAM basis \cite{Tscherbul:10,Tscherbul:12,Koyu:22}.  


\subsection{Ultracold atom-molecule collision dynamics in the TRAM basis}

 \begin{figure}[t!]
	\centering
	\includegraphics[width=1.05\columnwidth, trim = 0 0 0 10]{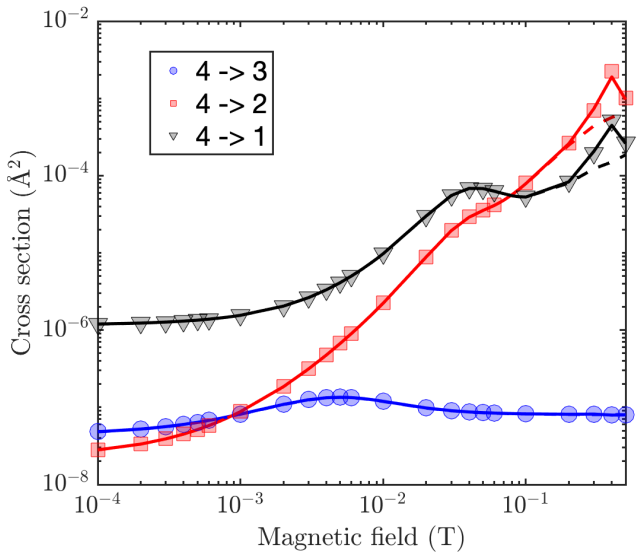}
	\caption{Magnetic field dependence of the integral cross sections for ultracold He~+~YbF collisions  with YbF molecules initially in the fully spin-polarized  state $\ket{4}$ (see Fig.~\ref{fig:energy_levels}) and the final states $\ket{3}$ (circles),  $\ket{2}$ (squares), and $\ket{1}$ (triangles).
		Solid lines -- TRAM calculations with $J_r^\text{max}=3$, dashed lines --   TRAM calculations with $J_r^\text{max}=2$, symbols -- benchmark calculations using the fully uncoupled basis \cite{Tscherbul:07}.  The collision energy is 1~mK.}
	\label{fig:xs_N0}
\end{figure}

To gauge the accuracy and computational efficiency of our TRAM basis, we have developed a new scattering code implementing CC equations in this  basis for $^2\Sigma$ molecule - $^1$S atom collisions as described in Sec.~IIA. The CC equations are parametrized by the matrix elements of the intramolecular Hamiltonian of YbF and of the He-YbF interaction potential as described in Ref.~\cite{Tscherbul:07} and in Sec. IIA. 

For comparison with previous quantum scattering calculations  on He~+~YbF \cite{Tscherbul:07} using the fully uncoupled basis, we used the same cutoff parameter for the rotational states $N^\text{max}=8$, but varied the cutoff parameter $J_r^\text{max}=2$ and 3. The CC equations are integrated using the log-derivative propagator \cite{Johnson:73,Manolopoulos:86} on a grid of $R$ extending from 3.84 to 80 $a_0$ with the grid step of 0.04~$a_0$. The following values of atomic and molecular masses were used in CC calculations (in atomic mass units): $m_\text{He}=3.01603$ and $m_\text{YbF}=192.9372652$. At the end of the propagation, the log-derivative matrix is transformed to the basis, in which the asymptotic Hamiltonian is diagonal, and then matched to the asymptotic boundary conditions to produce the reactance ($K$) and scattering ($S$) matrices, from which the integral cross sections are obtained using the standard expressions
\begin{equation}\label{ICS}
\sigma_{\gamma \to \gamma' }=\frac{\pi}{k_\gamma^2} \sum_M \sum_{l,l'} |\delta_{\gamma\gamma'}\delta_{ll'} -S^M_{\gamma l, \gamma'l'}|^2,
\end{equation}
where  the index $\gamma=1,2,\ldots$ labels the eigenvalues of the asymptotic Hamiltonian in the order of increasing energy  (see Fig.~\ref{fig:energy_levels}).  The unphysical states are assigned to the physical eigenstates that are closest in energy. The calculated cross sections are converged to $\leq 5$\%.

Figure~\ref{fig:xs_N0} compares state-to-state cross sections for ultracold He~+~YbF collisions  calculated using the TRAM basis set with reference calculations \cite{Tscherbul:07}. The  cross sections calculated using these completely unrelated bases are in excellent agreement with each other across the entire range of magnetic fields,  validating the accuracy of our TRAM calculations.  Taking the  $\ket{4}\to \ket{1}$  transition as an example, the relative differences between the TRAM cross sections  and the reference data do not exceed 0.1\% at magnetic fields  below 0.04 T (for $J_r^\text{max}=3$). The deviations increase to a few percent at higher magnetic fields, most likely due to incomplete convergence of the reference results \cite{Tscherbul:07}. 

Using a minimal basis set including the three lowest $J_r$ blocks ($J_r^\text{max}=2$) is sufficient for  magnetic fields below 0.1 T. Above this field value, a higher value of $J_r^\text{max}=3$ is required for the hyperfine transitions $\ket{4}\to\ket{1}$ and $\ket{4}\to\ket{2}$. As shown in Fig.~\ref{fig:energy_levels}, the final states $\ket{1}$ and $\ket{2}$ are low field-seeking, whereas the initial state $\ket{4}$ is high field-seeking.  Thus, the transitions  $\ket{4}\to\ket{1}$ and $\ket{4}\to\ket{2}$ release an increasing amount of energy  with increasing magnetic field, leading to the population of higher partial wave states in the outgoing collision channel, and necessitating the use of larger $J_r^\text{max}$.

The reference calculations employed the fully uncoupled basis set  with $N^\text{max}=8$ and $l^\text{max}=9$ \cite{Tscherbul:07}, leading to 2262 coupled channels for the total angular momentum projection $M=0$. By comparison, the TRAM basis with the same number of rotational states and $J_r^\text{max}=2$ includes 274 channels for the same value of $M$, a reduction in the number of coupled channels by the factor of 8.2.  The computational cost of solving CC equations scales as $N^3$ with the number of channels. Thus, using the TRAM basis leads to a 550-fold increase in computational efficiency compared to the fully uncoupled basis used in the previous calculations \cite{Tscherbul:07}.

\begin{figure}[t!]
	\centering
	\includegraphics[width=1.05\columnwidth, trim = 0 0 0 20]{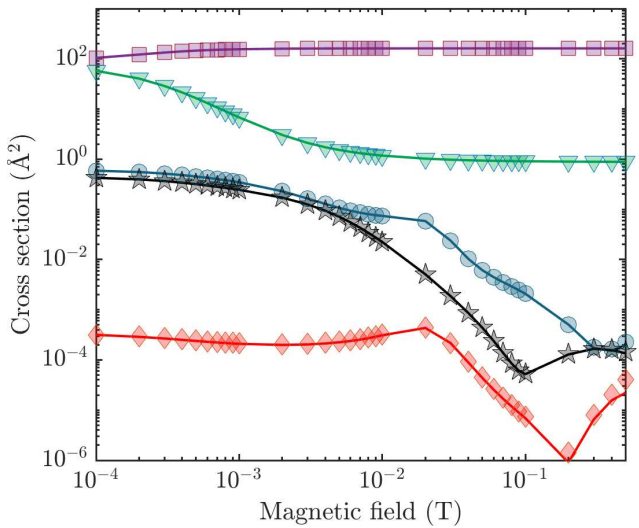}
	\caption{Magnetic field dependence of the integral cross sections for ultracold He~+~YbF collisions with YbF molecules initially in the rotationally excited state $\ket{16}$ and the  final Zeeman states $\ket{8}$ (circles), $\ket{4}$ (squares), $\ket{3}$ (triangles), $\ket{2}$ (diamonds), and $\ket{1}$ (stars) [see Fig.~\ref{fig:energy_levels}]. TRAM calculations with $J^\text{max}_r=3$ (solid lines) are compared with benchmark calculations  (symbols) using the fully uncoupled basis  \cite{Tscherbul:07}.  The collision energy is 1~mK.}
	\label{fig:xs_N1}
\end{figure}

Figure \ref{fig:xs_N1} shows the inelastic cross sections for YbF molecules initially in the highest low-field seeking hyperfine-Zeeman sublevel of the $N=1$ rotational state (state $\ket{16}$ in Fig.~\ref{fig:energy_levels}). Inelastic collisions with He atoms can either conserve or change the rotational state of YbF. All the state-to-state cross sections   calculated using the TRAM basis are in excellent agreement with the benchmark values, demonstrating the ability of the TRAM basis set to accurately describe ultracold collisions of rotationally excited molecules. The  unphysical states shown in the upper panel of Fig.~\ref{fig:energy_levels} do not affect the results of CC calculations for reasons discussed in Sec. IIIA.

\section{Summary and outlook}


\color{black}

Hyperfine interactions play an essential role in ultracold  atomic and molecular collisions, being largely responsible, for, e.g., positions and widths of magnetic Feshbach resonances in ultracold collisions of alkali-metal atoms \cite{Chin:10}.
Quantum scattering calculations must therefore account for hyperfine structure in order to provide a realistic picture  of   ultracold atom-molecule and molecule-molecule collision dynamics in the presence of external electromagnetic fields. 

While it is possible to include the hyperfine structure directly in the TAM representation, this approach is hindered by two difficulties. First,  unphysical states  show up in the  spectrum of threshold energy levels in the presence of external fields \cite{Tscherbul:10,Tscherbul:12,Koyu:22}. Even though these states do not affect the dynamics of ultracold collisions, and they can be eliminated by augmenting the basis set \cite{Koyu:22}, they can pose a challenge for bound-state calculations  \cite{Vexiau:19,Koyu:22}.
Second, constructing  the TAM basis functions requires multiple angular momentum coupling operations
 to form the eigenstates of $\hat{J}^2$ and $\hat{J}_Z$. This leads to complicated  expressions for the matrix elements of the interaction potential and/or the centrifugal kinetic energy \cite{Tscherbul:10,Suleimanov:12}, which can be challenging to implement in actual numerical calculations.



Here, we have explored an alternative basis set composed of products of eigenfunctions of the total  rotational angular momentum (TRAM) of the collision complex $J_r$ and the spin basis functions of its constituent atoms and molecules.  In the absence of spin-rotation interactions, $J_r$ is conserved and the scattering problem can be rigorously block-diagonalized  and solved separately for each value $J_r$ even in the presence of external magnetic field and isotropic hyperfine interactions.  This makes the TRAM basis particularly promising for molecules, whose collision dynamics is dominated by the isotropic hyperfine and Zeeman interactions (in addition to the electrostatic interaction potential). By contrast, the scattering Hamiltonian in the TAM representation \cite{Tscherbul:10,Tscherbul:12} is not block-diagonalizable in the presence of external magnetic fields. 

\textcolor{red}{In addition, the matrix elements of the Hamiltonian in the TRAM basis are simple to evaluate and program, and the unphysical states are eliminated  for $N=0$. Even though such states are still present for $N\ge 1$, their density is significantly reduced compared to that  in the TAM representation \cite{Koyu:22}. This is because the unphysical states in the TRAM basis arise due to the  matrix elements  of the anisotropic hyperfine and spin-rotation interactions that are off-diagonal in $J_r$ and independent of magnetic field. These matrix elements are small compared to those of the Zeeman interaction at moderate to strong magnetic fields (above $\simeq$100~G).
As a result, most of the eigenvalues of the asymptotic Hamiltonian in  the TRAM basis are close to the physical hyperfine-Zeeman states, and thus the density of unphysical states is low.
By contrast, in the TAM representation, the matrix elements off-diagonal in $J$ responsible for the unphysical states are due to the Zeeman interaction \cite{Tscherbul:10}, which is generally much stronger, and causes a substantially higher density  of unphysical states, especially for $N\ge 1$ \cite{Koyu:22}. }


We formulate the quantum scattering problem in the TRAM basis for   collisions of $^2\Sigma$ molecules with $^1$S atoms (Sec.~IIA) and with $^2$S atoms (Sec.~IIB). 
The generalized spin-rotation interactions, which couple the different values of  $J_r$, include the electron spin-rotation and anisotropic hyperfine, as well as the intermolecular magnetic dipole-dipole interaction (see Fig.~\ref{fig:cartoon}). 
 When these interactions are non-negligible, they can be incorporated in a straightforward manner by including several  $J_r$ states in the TRAM basis as discussed in Sec. II.
  This is  demonstrated by our CC calculations on ultracold $s$-wave He~+~YbF$(^2\Sigma)$ collisions in the regime, where the  electron spin-rotation and anisotropic hyperfine interactions in YbF are non-negligible (see Sec. III). With three lowest $J_r$ blocks in the basis, we observe excellent agreement of state-to-state scattering cross sections with prior calculations \cite{Tscherbul:07} using  8 times fewer TRAM basis functions, leading to a computational gain of about three orders of magnitude over the fully uncoupled basis  \cite{Tscherbul:07}.
  {Even larger  gains are expected for ultracold collisions of $^2\Sigma$ molecules with $^2$S atoms considered in Sec. IIB.}


Despite mounting experimental studies of Feshbach resonances in ultracold  atom-molecule  \cite{Yang:19,Wang:21,Zhen:22,Son:22,Zhen:22,Yang:22,Park:23b} and molecule-molecule \cite{Hu:19,Liu:22,Park:23} collisions, accurate quantum dynamics calculations on such collisions are currently beyond reach. As discussed above, this is partly due to the difficulties associated with including the effects of atomic and molecular hyperfine structure and/or external fields.  It is our hope that the computationally efficient and easy-to-implement TRAM representation will facilitate such calculations in the near future.
It would also be interesting to extend the TRAM approach  to ultracold molecule-molecule collisions and to other types of CC basis sets, such as those defined in the body-fixed coordinate frame \cite{Tscherbul:10,Morita:18,Morita:19c}.


\begin{acknowledgments}
We thank Dr. Masato Morita for verifying Eq.~\eqref{Vmdd_mel2}. 
The work at UNR was supported by the NSF through the CAREER program (grant No. PHY-2045681). The work at JILA was supported by the NSF (PHY-2012125) and by NASA/JPL (1502690).
\end{acknowledgments}

\appendix

\section{Matrix elements of the electron spin-rotation interaction in the TRAM basis}


Here, we derive the matrix elements of the intramolecular electron spin-rotation  interaction in the TRAM basis.
Expanding the spin-rotation interaction in rank-1 spherical tensor operators \textcolor{red}{$\hat{N}^{(1)}_{\pm1}=\mp \frac{1}{\sqrt{2}}\hat{N}_\pm = \mp \frac{1}{\sqrt{2}}(\hat{N}_X \pm \hat{N}_Y)$ and $\hat{N}^{(1)}_{0}=\hat{N}_Z$ (and similarly for $\hat{S}^{(1)}_{p}$) \cite{Zare:88}}
\begin{equation}\label{App11}
 \gamma_\text{sr} \hat{\mathbf{N}}\cdot \hat{\mathbf{S}} =  \gamma_\text{sr} \sum_{p=-1}^1 (-1)^q \hat{N}_p^{(1)} \hat{S}_{-p}^{(1)},
\end{equation}
and using the direct-product structure of the TRAM basis gives (omitting the basis functions $\ket{IM_I}$ for clarity)
\begin{multline}\label{App12}
\langle (Nl)J_rM_r | \langle S M_S | \gamma_\text{sr} \hat{\mathbf{N}}\cdot \hat{\mathbf{S}} 
 | (N'l')J_r'M_r' \rangle | SM_S'\rangle \\
=  \gamma_\text{sr} \sum_{p=-1}^1 (-1)^q \langle (Nl)J_rM_r | \hat{N}_p^{(1)} | (N'l')J_r'M_r' \rangle
\\ \times
 \langle S M_S | \hat{S}_{-p}^{(1)} | SM_S'\rangle
\end{multline}

Applying the Wigner-Eckart theorem to the rotational matrix element on the right-hand side 
\begin{multline}\label{App13}
 \langle (Nl)J_rM_r | \hat{N}_p^{(1)} | (N'l')J_r'M_r' \rangle
= (-1)^{J_r - M_r}
\\ \times
 \threejm{J_r}{-M_r}{1}{p}{J_r'}{M_r'}
 \langle (Nl)J_r || \hat{N}^{(1)} || (N'l')J_r' \rangle,
\end{multline}
and simplifying the resulting double-bar matrix element  \cite{Zare:88} we get
\begin{multline}\label{App14}
 \langle (Nl)J_rM_r | \hat{N}_p^{(1)} | (N'l')J_r'M_r' \rangle
= (-1)^{J_r - M_r}
\\ \times
 \threejm{J_r}{-M_r}{1}{p}{J_r'}{M_r'}
\delta_{ll'}(-1)^{N+l+J_r'+1}[(2J_r+1)(2J_r'+1)]^{1/2}
\\ \times
\sixj{N}{J_r'}{J_r}{N}{l}{1} [(2N+1)N(N+1)]^{1/2}\delta_{NN'}
\end{multline}
Using
\begin{multline}\label{App15}
 \langle S M_S | \hat{S}_{-p}^{(1)} | SM_S'\rangle
 = (-1)^{S-M_S} \threejm{S}{-M_S}{1}{-p}{S}{M_S'}
 \\ \times [(2S+1)S(S+1)]^{1/2}
\end{multline}
in combination with Eqs.~\eqref{App14} and \eqref{App12}, we obtain Eq.~\eqref{mel_Hmol_term3sr} of the main text.


\section{Matrix elements of the anisotropic hyperfine interaction in the TRAM basis}

The anisotropic hyperfine interaction given by the last term in Eq.~\eqref{Hmol_spin_rot}  is composed of three tensor operators, which depend on the rotational, electron spin, and nuclear spin variables
\begin{multline}\label{App21}
\hat{H}_{\textrm{mol}}^{ahf}= 
\frac{c\sqrt{6}}{3}\biggl{(}\frac{4\pi}{5}\biggr{)}^{1/2}\sum_{q=-2}^2 (-1)^qY_{2-q}(\theta_r,\phi_r)
          [\hat{\mathbf{I}}\otimes \hat{\mathbf{S}}]^{(2)}_q,
\end{multline}

Taking the matrix elements of this expression in the TRAM  basis, we find
 \begin{multline}\label{App22}
\langle (Nl)J_rM_r | \langle S M_S | \langle IM_I | \hat{H}_\text{mol}^\text{ahf} | (N'l')J_r'M_r' \rangle | SM_S'\rangle |IM_I'\rangle 
\\
=  \frac{c\sqrt{6}}{3}\biggl{(}\frac{4\pi}{5}\biggr{)}^{1/2}\sum_{q=-2}^2 (-1)^q
\langle (Nl)J_rM_r | Y_{2-q} | (N'l')J_r'M_r' \rangle
 \\ \times
 \langle S M_S | \langle IM_I |  [\hat{\mathbf{I}}\otimes \hat{\mathbf{S}}]^{(2)}_q | SM_S'\rangle |IM_I'\rangle
\end{multline}
The first matrix element on the right-hand side can be evaluated  using the Wigner-Eckart theorem \cite{Zare:88}
 \begin{multline}\label{App23}
\langle (Nl)J_rM_r | Y_{2-q} | (N'l')J_r'M_r' \rangle = (-1)^{J_r-M_r}
\\ \times 
 \threejm{J_r}{-M_r}{2}{-q}{J_r'}{M_r'}
 \langle (Nl)J_r || Y^{(2)} || (N'l')J_r' \rangle
\end{multline}
with the double-bar matrix element given by \cite{Zare:88}
 \begin{multline}\label{App23}
 \langle (Nl)J_r || Y^{(2)} || (N'l')J_r' \rangle = \delta_{ll'}(-1)^{l+J_r'} 
\\ \times 
[(2J_r+1)(2J_r'+1)]^{1/2}
\sixj{N}{J_r'}{J_r}{N'}{l}{2} \left(\frac{5}{4\pi}\right)^{1/2}
\\ \times
[(2N+1)(2N'+1)]^{1/2}
 \threejm{N}{0}{2}{0}{N'}{0}
\end{multline}

It remains to consider the spin matrix element
 \begin{multline}\label{App24}
 \langle S M_S | \langle IM_I |  [\hat{\mathbf{I}}\otimes \hat{\mathbf{S}}]^{(2)}_q | SM_S'\rangle |IM_I'\rangle = \sum_{q_I,q_S}(-1)^q \sqrt{5}
 \\  \times
  \threejm{1}{q_I}{1}{q_S}{2}{-q} \langle I M_I | \hat{I}^{(1)}_{q_I}   | IM_I'\rangle  \langle SM_S | \hat{S}^{(1)}_{q_S}|SM_S'\rangle
\end{multline}
where we have used the definition of the tensor product of two spherical tensor operators \cite{Zare:88}. Using once again the Wigner-Eckart theorem
 \begin{multline}\label{App25}
 \langle IM_I |  \hat{I}^{(1)}_{q_I}  | IM_I'\rangle = (-1)^{I-M_I}  \threejm{I}{-M_I}{1}{q_I}{I}{M_I'}
 \\ \times
 [(2I+1)I(I+1)]^{1/2}
\end{multline}
and the corresponding expression for the matrix elements of $\hat{S}^{(1)}_{q_S}$, we obtain
\begin{multline}\label{App26}
 \langle S M_S | \langle IM_I |  [\hat{\mathbf{I}}\otimes \hat{\mathbf{S}}]^{(2)}_q | SM_S'\rangle |IM_I'\rangle = \sum_{q_I,q_S}(-1)^q \sqrt{5}
 \\  \times
  \threejm{1}{q_I}{1}{q_S}{2}{-q} 
  (-1)^{I-M_I+S-M_S}  \threejm{I}{-M_I}{1}{q_I}{I}{M_I'}  
  \\ \times [(2I+1)I(I+1)(2S+1)S(S+1)]^{1/2}
   \threejm{S}{-M_S}{1}{q_S}{S}{M_S'} 
\end{multline}
This expression, combined with Eqs.~\eqref{App23} and \eqref{App22}, gives Eq.~\eqref{mel_Hmol_term3ahf} of the main text.

\section{Matrix elements of the magnetic dipole-dipole interaction in the TRAM basis}

Finally, we consider the matrix elements of the magnetic dipole-dipole interaction, which does not affect the nuclear spin degrees of freedom, and is thus diagonal in $M_I$ and $M_{I_a}$. Omitting the nuclear spin basis functions and using the shorthand notation $ |M_S M_{S_a}\rangle$ for the electron spin basis functions $ | SM_S \rangle |S_a M_{S_a}\rangle$ the matrix elements take the form
\begin{multline}\label{AppC0}
 \langle (Nl)J_rM_r | \langle M_S M_{S_a} | \hat{V}_\text{mdd} |(N'l')J_r'M_r' \rangle | M_S' M_{S_a}'\rangle   
 \\  =  - \sqrt{\frac{24\pi}{5}} \frac{\alpha^2}{R^3} \sum_{q} (-1)^q 
  \langle (Nl)J_rM_r | Y_{2,-q}(\hat{R})|(N'l')J_r'M_r' \rangle 
  \\ \times
 \langle M_S M_{S_a} | [\hat{\mathbf{S}}\otimes\hat{\mathbf{S}}_a]^{(2)}_q   | M_S' M_{S_a}'\rangle   
\end{multline}

The Wigner-Eckart theorem allows us to factorize the orbital matrix element involving the spherical harmonics $Y_{2,-q}(\hat{R})$  as
\begin{multline}\label{AppC1}
 \langle (Nl)J_rM_r | Y_{2,-q}(\hat{R}) | (N'l')J_r'M_r' \rangle
= (-1)^{J_r - M_r}
\\ \times
 \threejm{J_r}{-M_r}{2}{-q}{J_r'}{M_r'}
 \langle (Nl)J_r || \hat{Y}^{(2)} || (N'l')J_r' \rangle.
\end{multline}
Evaluating the  double-bar matrix element  \cite{Zare:88} we obtain
\begin{multline}\label{AppC2}
 \langle (Nl)J_rM_r | Y_{2,-q}(\hat{R}) | (N'l')J_r'M_r' \rangle
= (-1)^{J_r - M_r}
\\ \times
 \threejm{J_r}{-M_r}{2}{-q}{J_r'}{M_r'}
\delta_{NN'}(-1)^{N+l'+J_r}[(2J_r+1)(2J_r'+1)]^{1/2}
\\ \times
\sixj{l}{J_r'}{J_r}{l'}{N}{2} (-1)^l [(2l+1)(2l'+1)]^{1/2}
\sqrt{\frac{5}{4\pi}} 
 \threejm{l}{0}{2}{0}{l'}{0}
\end{multline}

The spin matrix element in Eq.~\eqref{AppC0} can be evaluated as described in Appendix B
\begin{multline}\label{AppC3}
 \langle M_S M_{S_a} |  [\hat{\mathbf{S}}\otimes \hat{\mathbf{S}}_a]^{(2)}_q | M_S' M_{S_a}' \rangle = \sum_{q_S,q_{S_a}}(-1)^q \sqrt{5}
 \\  \times
  \threejm{1}{q_S}{1}{q_{S_a}}{2}{-q} 
  (-1)^{S-M_S+S_a-M_{S_a}}  \threejm{S}{-M_S}{1}{q_S}{S}{M_S'}  
  \\ \times [(2S+1)S(S+1)]^{1/2}[(2S_a+1)S_a(S_a+1)]^{1/2} 
  \\ \times
   \threejm{S_a}{-M_{S_a}}{1}{q_{S_a}}{S_a}{M_{S_a}'} 
\end{multline}
Combining this result with Eq.~\eqref{AppC2} and setting $q_S = M_S-M_S'$, $q_{S_a} = M_{S_a} - M_{S_a}'$, and $-q=M_r - M_r'$, we obtain   Eq.~\eqref{Vmdd_mel2}  of the main text.

\end{document}